\documentclass[12pt]{article}
\usepackage{latexsym,cite}
\input amssym.def
\input amssym.tex

\makeatletter

\@addtoreset{equation}{section}
\makeatother

\newcommand{\diag}{{\rm diag}}

\newcommand{\half}{{{\textstyle\frac{1}{2}}}}

\newcommand{\be}{\begin{equation} }
\newcommand{\ee}{\end{equation} }
\newcommand{\ba}{\begin{array}}
\newcommand{\ea}{\end{array}}

\newcommand{\so}{\mathbf{so}}

\newcommand{\osp}{\mathbf{osp}}

\def\F{{\cal{F}}}
\def\G{{\cal{G}}}

\def\M{{\cal M}}
\def\K{{\cal{K}}}
\def\X{{\cal{X}}}

\newcommand{\Q}{{\cal{Q}}}
\def\R{{\mathbf R}}

\def\RAdS{{R_{\scriptscriptstyle{{AdS}}}}}

\def\QX{{Q_{{\scriptscriptstyle X}}}}
\def\QP{{Q_{{\scriptscriptstyle P}}}}

\def\HU1{{H_{{\scriptscriptstyle{\rm u}(1)}}}}
\def\HSU{{H_{{\scriptscriptstyle{\rm su}(N)}}}}
\def\VSU{{V_{{\scriptscriptstyle{\rm su}(N)}}}}

\def\tr{{\rm tr}}
\def\trSU{{\rm tr}_{\scriptscriptstyle{{\rm su}(N)}}}
\def\trU1{{\rm tr}_{\scriptscriptstyle{{\rm u}(1)}}}
\def\Tr{{\rm Tr}}

\def\I_M{{I_{\scriptscriptstyle M\times M}}}

\def\YM{{\scriptscriptstyle {\rm YM}}}
\def\6D{{\scriptscriptstyle {6D}}}
\def\4D{{\scriptscriptstyle {4D}}}
\def\3D{{\scriptscriptstyle {3D}}}
\def\2Dstring{{\scriptstyle {2D\,{\rm string}}}}
\def\3DM{{\scriptstyle {3D{\cal M}{\rm theory}}}}

\def\N{{\cal  N}}

\def\Cosp{{\cal C}_{\osp{\scriptstyle{(1|2,R)}}}}
\def\Cso{{\cal C}_{\so{\scriptstyle{(1,2)}}}}

\def\A{{A_{0}}}

\def\cL{{{\cal L}}}

\def\dis{\displaystyle}

\begin{document}
\begin{titlepage}
\title{\vskip -60pt
{\small
\begin{flushright}
hep-th/yymmnnn
\end{flushright}}
\vskip 20pt
Noncritical  $\osp(1|2,\R)$ ${\M}$-theory matrix model with\\ an arbitrary
time      dependent cosmological constant \\ ~~~\\~~~ }
\author{Jeong-Hyuck Park\thanks{E-mail address:\,\,park@mppmu.mpg.de}}
\date{}
\maketitle
\vspace{-1.0cm}
\begin{center}
~~~\\
\textit{Max-Planck-Institut fur Physik,  Fohringer Ring 6,  80805  Munchen,  Germany}\\
~~~\\
~~~\\
~~~\\
\end{center}
\begin{abstract}
\noindent Dimensional reduction of the ${D=2}$ minimal super
Yang-Mills to the   ${D=1}$ matrix quantum mechanics  is shown to
double the number of dynamical supersymmetries,  from  $\N=1$ to
$\N=2$.  We analyze  the most general supersymmetric deformations of
the latter, in order to construct  the noncritical $3D$ $\M$-theory
matrix model on \textit{generic} supersymmetric  backgrounds. It
amounts  to adding quadratic and linear  potentials with  arbitrary
time dependent coefficients, namely,  a cosmological `constant,'
$\Lambda(t)$, and an  electric  flux background, $\rho(t)$,
respectively.  The resulting matrix model enjoys,  irrespective of
$\Lambda(t)$ and $\rho(t)$,    two dynamical supersymmetries which
further  reveal three hidden $\so(1,2)$ symmetries. All together
they  form the supersymmetry algebra, $\osp(1|2,\R)$.  Each
$\so(1,2)$ multiplet  in the Hilbert space  visualizes  a dynamics
constrained on either Euclidean or Minkowskian  $dS_{2}/AdS_{2}$
space,  depending on its Casimir. In particular, all the unitary
multiplets  have   the  Euclidean $dS_{2}/AdS_{2}$ geometry.    We
conjecture that the  matrix model provides holographic duals  to the
$2D$ superstring theories on various backgrounds having   the
spacetime signature Minkowskian if $\Lambda(t)>0$,  or Euclidean if
$\Lambda(t)<0$.
\end{abstract}
~\\
{\small
\begin{flushleft}
\end{flushleft}}
\thispagestyle{empty}
\end{titlepage}
\newpage


\section{Introduction}
String or $\M$-theory dress    all the known supersymmetric gauge
theories with  the insightful geometrical  pictures  by the notion
of holography or $AdS$/CFT
correspondence~\cite{Maldacena:1997re,Aharony:1999ti}. In
particular, the symmetry group of a  gauge theory is identified as
the isometry of   the corresponding higher dimensional
string/$\M$-theory   background. Conversely, different  string
theories  - bosonic or supersymmetric, critical or noncritical - on
various   backgrounds
  are expected to have   holographic dual   gauge theories.   \\

However, despite of some
progress~\cite{Marinari:1990jc,Jevicki:1993zg,McGreevy:2003dn,Dabholkar:1991te,Gukov:2003yp,
Takayanagi:2004ge,Verlinde:2004gt,CollectiveField},  the conformal
dual description of the noncritical $2D$ superstring is a yet
unresolved problem.   In the present paper, we attempt to address
the issue from the $\M$-theory   point of
view~\cite{Townsend:1998qp,Horava:2005tt,McGuigan:2004sq}. The
spacetime dimension two is singular in the sense that  the
holographic dual of $2D$ superstring theory, whatever its concrete
form is, should share many common features with  the corresponding
noncritical
 $\M$-theory matrix model.\\

As is well known, superstring lives in $2$, $3$, $4$, $6$ and $10$
dimensions,  while the supermembrane exits in   dimensions one
higher, i.e. $3$, $4$, $5$, $7$ and $11$,  since only in those
spacetime dimensions  the relevant Fierz  identities hold.  Although
the pioneering work on super $p$-branes~\cite{Achucarro:1987nc}
excludes the possibility of the space-filling $p$-branes i.e.
$p$-branes   propagating in $(p+1)$-dimensional target spacetime,
supermembrane \textit{does} exit in three dimensions,  since  the
Fierz identity for the supermembrane works manifestly, from
$\gamma^{012}=1$,
\be
\dis{({\rm d}\bar{\theta}\gamma_{\mu}{\rm
d}\theta)({\rm d}\bar{\theta}\gamma^{\mu\nu}{\rm
d}\theta)=\epsilon^{\mu\nu\lambda}({\rm
d}\bar{\theta}\gamma_{\mu}{\rm d}\theta)({\rm
d}\bar{\theta}\gamma_{\lambda}{\rm d}\theta)=0\,,}
\ee
where ${\rm d}\theta$ is a  bosonic spinor.
The matrix regularization~\cite{deWit:1988ig,Hoppe:2002km} of
the  supermembrane  prescribes the  replacement of  the Poisson bracket
appearing in the light cone gauged membrane action by a matrix commutator.
For $3D$ supermembrane action, it  leads to a \textit{supersymmetric} and
\textit{gauged} version of a \textit{one}  matrix model, where the
local gauge symmetry originates from the area preserving diffeomorphism for
the Poisson bracket, and the appearance of only one matrix is due
to the light cone gauge, i.e. $3-2=1$. \\

The resulting  $N\times N$  matrix model,
at least for the `flat' $3D$ background, can be also obtained by the dimensional reduction
of the $2D$ minimal super Yang-Mills\footnote{Recently all the minimal noncritical
 super Yang-Mills (except ${D=3}$) have been  identified  in
 the noncritical superstring theories~\cite{Ashok:2005py}.}
 to $D=1$, and it is  supposed to  describe exactly
 the  D0-brane dynamics of  the discrete light cone momentum sector, $p_{-}=N/R$,
 in $\M$-theory compactified on   a light-like circle, $x^{-}\sim x^{-}+2\pi R$,
 as initially proposed by  Banks, Fischler, Shenker and  Susskind
 for the critical $\M$-theory~\cite{Banks:1996vh,Susskind:1997cw}.
 As for the D0-branes, the local gauge symmetry is required to reflect
 the identical nature of the $N$ D-particles~\cite{Park:2002eu}.\\

Also for the noncritical $2D$ superstring,
almost by definition, its holographic dual
should be \textit{one dimensional, supersymmetric} and \textit{gauged  theories}.
In the presence of RR electric field, $F$,
the low energy effective action of  $2D$  string theory   typically reads,
neglecting the massless tachyon and putting
$\alpha^{\prime}\equiv 1$ \cite{Douglas:2003up,Thompson:2003fz,Strominger:2003tm},
\be
\dis{S_{\,{2D}}=\int{\rm d}^{2}x\sqrt{-g}\left[e^{-2\Phi}\left(
8+R+4\left(\nabla\Phi\right)^{2}\right)-\half F^{2}\right]\,,}
\label{S2D}
\ee
where $-\half F^{2}$ plays the role of the negative cosmological constant,
and the solutions are characterized by
the $AdS_{2}$-like geometries.\footnote{In two dimensions the geometries
of $AdS_{2}$ and $dS_{2}$ coincide, and   we will distinguish them by the sign of the
`cosmological constant'.   Also it is to be reminded that
\[
\ba{ll}
k_{0}^{2}-k_{1}^{2}-k_{2}^{2}=R^{2}>0~~~&~~~\mbox{:~Euclidean~}AdS_{2}/dS_{2}~~\mbox{
(hyperboloid~of~two~sheets)}\,,\\
{}\\
k_{0}^{2}-k_{1}^{2}-k_{2}^{2}=-R^{2}<0~~~&~~~\mbox{:~Minkowskian~}AdS_{2}/dS_{2}~~
\mbox{(hyperboloid~of~one~sheet)}\,.
\ea
\]}
Indeed,  switching  off the dilaton  completely we have the $AdS_{2}$ solution, while
turning on $\Phi$, one has
static extremal black hole-like solutions~\cite{Banks:1992xs,Berkovits:2001tg,Gukov:2003yp}.
In the asymptotic region the latter becomes the usual linear dilaton vacuum,
and in the ``near-horizon" region it approaches  to $AdS_{2}$  with the
dilaton  reaching the critical value, $\Phi_{c}=-\ln(\textstyle{\frac{1}{4}}F)$.\\

However, the effective action, (\ref{S2D}),  can not be thoroughly
trusted due to the $\alpha^{\prime}$ corrections as well as the tachyon tadpoles.
Necessarily one has  to work  on  the full sigma model (e.g. \cite{Kazama:1988qp})
with the difficulty of dealing with  background fluxes.  Hence to find the exact nontrivial
superstring  background is not an easy task. And also for the $\M$-theory,
the matrix regularization of the supermembrane action is not always straightforward
for generic nontrivial backgrounds. \\

In this work, we take  \textit{supersymmetry} itself
as the principal  guideline  to tackle the problem of  constructing
the noncritical $3D$ $\M$-theory matrix  model on generic supersymmetric backgrounds.
Namely  after the dimensional reduction of the ${D=2}$ super Yang-Mills to the  $D=1$
 matrix quantum mechanics,  we analyze all  the possible  deformations of the latter
 without breaking  any supersymmetry.  We show that the most general supersymmetric
 deformations simply amount to adding  quadratic and linear  potentials with  arbitrary
  time dependent coefficients, namely,  a  cosmological `constant,'  $\Lambda(t)$,
  and an electric  flux background, $\rho(t)$, respectively. The latter  couples to
  the $\mathbf{u}(1)$ sector or the ``center of mass"  only, while
  the $\mathbf{su}(N)$ sector and the $\mathbf{u}(1)$ sector are completely decoupled.
  Remarkably  we find that, irrespective of  $\Lambda(t)$ and $\rho(t)$,  the resulting
  matrix model always enjoys two dynamical supersymmetries, not just  one  as in the $2D$
  minimal super Yang-Mills. Namely  after the dimensional reduction, the number of
  supersymmetry is doubled,  from $\N=1$ to $\N=2$.  Furthermore, again for arbitrary
  $\Lambda(t)$ and $\rho(t)$, these two  supersymmetries reveal three hidden nontrivial
  bosonic symmetries. All together the five symmetries  form the super Lie algebra,
  $\osp(1|2,\R)$, where the even part corresponds to $\so(1,2)\,$  i.e. the isometry
  of the Euclidean or Minkowskian ${dS}_{2}/{AdS}_{2}$.   We introduce   a
  projection map from the phase space   to a three dimensional `$\so(1,2)$ hyperspace'
  associated with  the  bosonic symmetries.  The dynamics therein   is always constrained on
  a two dimensional rigid surface, Euclidean  $dS_{2}/AdS_{2}$ or  Minkowskian  $dS_{2}/AdS_{2}$,
   depending on the sign of the   $\so(1,2)$ Casimir  for  each
   multiplet  in the Hilbert  space. The richness of the  matrix model comes from
   the arbitrariness of the time dependent coefficients, ${\Lambda}(t)$,  $\rho(t)$,
   and the vast amount of  supermultiplets existing  in the Hilbert  space each of which has
   its own two dimensional geometries. \\

The organization of the present paper is as follows.
In section \ref{MATRIX-MODEL}, we analyze
  the  most general supersymmetric deformations of the   matrix model having  the $2D$
  super Yang-Mills origin. We discuss its symmetries, Hamiltonian dynamics and the BPS
  configurations. We also comment on the relation to the matrix cosmology.
Section \ref{SecOSP} is devoted to the  detailed  analysis on the underlying supersymmetry
algebra,   $\osp(1|2,\R)$,  both from the kinematical and dynamical  point of view.
In particular, we show that all the  unitary  multiplets  correspond  to the
Euclidean ${dS_{2}/AdS_{2}}$ geometry, rather than the Minkowskian one.
The last section, Sec.\ref{CONCLUSION}  contains     our  conjecture   that
the matrix model with different choices of $\Lambda(t)$ and $\rho(t)$ may
provide holographic duals to various $2D$ superstring or superconformal theories.  \newpage

\section{Noncritical $\osp(1|2,\R)$ $\M$-theory matrix model\label{MATRIX-MODEL}}
\subsection{Derivation of the matrix model and SUSY  enhancement}
In two dimensional Minkowskian spacetime the fermion  satisfies the Majorana-Weyl condition, resulting in only one component real spinor.  After the dimensional reduction to $D=1$, the $2D$ super Yang-Mills leads to the following supersymmetric  matrix model,  which can be also obtained by the matrix regularization of the $3D$ supermembrane action in the light cone gauge,
\be
{\cL}=\tr\Big[\half D_{t}XD_{t}X+i\half\psi D_{t}\psi
+X\psi\psi\Big]\,,
\label{2D0}
\ee
where $X$, $\psi$ are respectively bosonic or fermionic $N\times N$ Hermitian matrices. With a gauge potential, $\A=\A^{\dagger}$, the covariant time derivative reads,  in our convention,
\be
D_{t}=\partial_{t}\,-i[\A\,,\,\,~\,\,]\,.
\ee
Bosons, $X,\A$, have the mass dimension 1, while the fermion, $\psi$, has the mass dimension  $\frac{3}{2}$,  so that the Lagrangian has the mass dimension, $4$. \\

The supersymmetry transformation, $\delta_{\YM}$,  descending from the $2D$ super Yang-Mills theory  is,  with a constant supersymmetry parameter, $\varepsilon$,
\be
\ba{ll}
\delta_{\YM} \A=\delta_{\YM}X=i\psi\varepsilon\,, ~~~~~&~~~~~
\delta_{\YM}\psi=D_{t}X\varepsilon\,.
\ea
\label{susy20}
\ee
~\\

Now we look for the generalizations  of the above Lagrangian as well as the supersymmetry transformations.
 First of all, we note from
\be
\tr\Big[i\half\psi D_{t}\psi
+X\psi\psi\Big]=\tr\Big[i\half\psi\partial_{t}\psi+(X-\A)\psi\psi\Big]\,,
\ee
that in order to cancel the cubic terms of  $\psi$ in any possible supersymmetry variation which will transform the bosons, $(X-\A)$  to the fermion,  it is inevitable   to impose\footnote{Essentially this rigidity corresponds to the Fierz identity,
$\tr\!\left(\bar{\psi}\gamma^{\mu}[\bar{\varepsilon}\gamma_{\mu}\psi,\psi]\right)=0\,,$ relevant to  the existence of the  minimal super Yang-Mills in $2$, $3$, $4$, $6$, $10$  dimensions.}
\be
\delta \A=\delta X\,.
\ee
Hence,  introducing   a time      dependent function, $f(t)$,  we let the generalized supersymmetry transformation be
\be
\ba{ll}
\delta \A=\delta X=if(t)\psi\varepsilon\,, ~~~~~&~~~~~
\delta\psi=\Big(f(t)D_{t}X+\Delta\Big)\varepsilon\,,
\ea
\label{susy2ans}
\ee
where $\Delta$ is a bosonic quantity having the mass dimension $2$, and its  explicit form  is to be determined shortly.  After some straightforward manipulation,  we obtain
\be
\delta{\cL}=\tr\!\left[
i\psi\varepsilon\left(D_{t}\Big(\dot{f}X+\Delta\Big)-\ddot{f}X+i[X,\Delta]\right)
\right]~+~\partial_{t}\K\,,
\ee
where the total  derivative term is given by
\be
\K=\tr\!\left(D_{t}X\delta X-i\half\psi\delta\psi\right)\,.
\ee
Of course, the simplest case where  $f(t)=1$ and $\Delta=0$ reduces to the supersymmetry of the original  $2D$ super Yang-Mills, (\ref{susy20}).    For the generic cases,   we are obliged to  set
\be
\Delta=-\dot{f}X-\kappa 1\,,
\label{Deltafix}
\ee
and obtain the following supersymmetry invariance,
\be
\delta\left[{\cL}+
\tr\!\left(\half{(\ddot{f}/f)}X^{2}+(\dot{\kappa}/f)X\right)\right]=\partial_{t}\K\,.
\ee
This   essentially leads to a  novel  supersymmetric matrix model with
two arbitrary time dependent functions, $\kappa(t)$, $f(t)$, as spelled out
in Eq.(\ref{FINAL2}).\\

For  given functions, ${\Lambda}(t)$ and $\rho(t)$, there exit two  sets of `solutions' given by   $f(t)$, $\kappa(t)$ to satisfy   the following second order differential equations,\footnote{The integral constant for $\kappa(t)$ corresponds to the kinematical supersymmetry.}
\be
\ba{ll}
\dis{\Lambda={\ddot{f}/f}\,,}~~~~~&~~~~~\dis{
\rho={\dot{\kappa}/f}\,.}
\ea
\ee
Thus, surprisingly,  there are two dynamical supersymmetries in the  matrix model, even for
the case,\footnote{This kind of  supersymmetry enhancement after the dimensional reduction can
 be  also noticed  elsewhere. For example, in the earlier works~\cite{Hyun:2002fk,Hyun:2003se},
 we derived the effective worldvolume  gauge theories for the longitudinal D5 and D2 branes on
 the maximally supersymmetric $11D$  $pp$-wave background. After the dimensional reductions to
   ${D=1}$,  both of them lead to a  matrix quantum mechanics  which is  equivalent to the BMN
   $\M$-theory matrix model~\cite{Berenstein:2002jq} up to  field redefinitions. The formers have
   only four  dynamical supersymmetries, while the BMN model has $32$  supersymmetries, $16$
   dynamical and $16$ kinematical.   The physical reason for the enhancement is that
   the D-branes  which  the higher dimensional gauge theories describe  preserve only the
   fraction of the full $\M$-theory  supersymmetries, $\frac{4}{32}$.  The same reasoning
   also holds for the present  $\osp(1|2,\R)$ $\M$-theory matrix model having  three
   supersymmetries, two  dynamical  and one kinematical.
 As we see shortly,  all the BPS states preserve  only one  supersymmetry,
 breaking the other two.  Hence, the minimal $2D$ super Yang-Mills can be interpreted
 as the worldvolume action of the longitudinal M2-brane   which preserves only
 one supersymmetry. However, it remains somewhat mysterious that the total number
 of supersymmetries is three, a rather unusual odd number.}  ${\Lambda}=\rho=0$.
 This will  further reveal  three   nontrivial  bosonic  symmetries as we discuss
 in the next subsection. \\

Rather than taking (\ref{Deltafix}) one might attempt to close the supersymmetry invariance
by adding other terms to ${\cL}$. However, since  there exits only one component spinor,
there can not be any mass term for the fermion\footnote{This is a special feature  only present
in the matrix quantum mechanics  of the $2D$ super Yang-Mills origin.  In fact, in the higher
dimensional cases  one needs to add the fermion mass term  for the supersymmetry invariance
as in the BMN matrix model~\cite{Berenstein:2002jq} or \cite{future}.}  as $\tr(\psi\psi)=0$.
Thus, as long as we restrict on the `non-derivative corrections', the above generalization is
the most generic one. \\
~\\

\subsection{Noncritical $\osp(1|2,\R)$ $\M$-theory matrix model  : Final form}
With an arbitrary time      dependent `cosmological constant',  $\Lambda(t)$,    having  the mass dimension two and an arbitrary time dependent  `electric flux background', $\rho(t)$, having the mass dimension three,  the generic form of the  noncritical $3D$ $\M$-theory matrix model  reads
\be
{\cal L}_{{\mathbf{osp}}(1|2,\R)}=\tr\Big[\half\left(D_{t}X\right)^{2}+i\half\psi D_{t}\psi
+X\psi\psi +\half {\Lambda}(t)X^{2}+\rho(t)X\Big]\,.
\label{FINAL2}
\ee
The Lagrangian  corresponds to  the most general supersymmetric deformations of  the  `$\N=2$'  matrix quantum mechanics   of the $2D$ super Yang-Mills  origin.  The  matrix model is to describe   the noncritical $3D$ supermembrane in a controllable manner through  the  matrix regularization, and our claim is further that it  also provides    holographic duals   to  $2D$  superstring theories, as discussed in the last section.\\

The matrix model is   equipped with the standard local gauge symmetry,
\be
\ba{llll}
X~\longrightarrow~ gXg^{-1}\,,~~&~~\psi~\longrightarrow~ g\psi g^{-1}\,,~~&~~\A~\longrightarrow~ g\A g^{-1}-i\partial_{t}gg^{-1}\,,~~&~~g\in \mathbf{U}(N)\,,
\ea
\label{gaugesym}
\ee
and enjoys \textit{two} dynamical supersymmetries,
\be
\ba{ll}
\delta_{\pm} \A=\delta_{\pm} X=if_{\pm}(t)\psi\varepsilon_{\pm}\,, ~~~~~&~~~~~
\delta_{\pm}\psi=\Big(f_{\pm}(t)D_{t}X-\dot{f}_{\pm}(t)X-\kappa_{\pm}(t)
1\Big)\varepsilon_{\pm}\,,
\ea
\label{SUSYf2}
\ee
where  $\varepsilon_{+}$, $\varepsilon_{-}$, are two real supersymmetry parameters, and
$f_{\pm}(t)$, $\kappa_{\pm}(t)$  are the two different solutions of the  second order differential equations,
\be
\ba{ll}
\dis{\ddot{f}_{\pm}(t)=f_{\pm}(t){\Lambda}(t)\,,}~~~~~&~~~~~
\dis{{\kappa}_{\pm}(t):=\int^{t}_{t_{0}}\!{\rm d}t^{\prime}\,\rho(t^{\prime})f_{\pm}(t^{\prime})\,.}
\ea
\label{Raychaudhuri}
\ee
The above  two dynamical supersymmetries further reveal three hidden nontrivial
bosonic symmetries,\footnote{
It is worth to note  that the three bosonic symmetries are still valid in the bosonic matrix model obtained after putting $\psi\equiv 0$,
\[
{\cal L}_{{\mathbf{so}}(1,2)}=\tr\Big[\half\left(D_{t}X\right)^{2} +\half {\Lambda}(t)X^{2}+\rho(t)X\Big]\,.
\label{BosonicM}\]} which we denote by $~\delta_{\scriptscriptstyle{++}}\,$, $~\delta_{\scriptscriptstyle{{--}}}\,$, $~\delta_{\scriptscriptstyle{{\{+,-\}}}}\,$,  {~}in order to indicate the anti-commutator origin   from the  two supersymmetries,
\be
\ba{l}
\delta_{\scriptscriptstyle{++}} \A=\delta_{\scriptscriptstyle{++}} X=f_{+}\left(f_{+}
D_{t}X-\dot{f}_{+}X-\kappa_{+}1\right)\,,~~~~~~~
\delta_{\scriptscriptstyle{++}}\psi=0\,,\\
{}\\
\delta_{\scriptscriptstyle{--}} \A=\delta_{\scriptscriptstyle{--}} X=f_{-}\left(f_{-}
D_{t}X-\dot{f}_{-}X-\kappa_{-}1\right)\,,~~~~~~~
\delta_{\scriptscriptstyle{--}}\psi=0\,,\\
{}\\
\delta_{\scriptscriptstyle{{\{+,-\}}}} \A=\delta_{\scriptscriptstyle{{\{+,-\}}}} X=2f_{+}f_{-}D_{t}X-\left(f_{+}\dot{f}_{-}+f_{-}\dot{f}_{+}\right)X-
\left(f_{+}\kappa_{-}+f_{-}\kappa_{+}\right)1\,,~~
\delta_{\scriptscriptstyle{{\{+,-\}}}}\psi=0\,.
\ea
\label{hidden}
\ee
Since $\frac{{\rm d}~}{{\rm d}t}\left(f_{+}\dot{f}_{-}-f_{-}\dot{f}_{+}\right)=0$,
if we set a non-zero constant,
\be
\displaystyle{c:=f_{+}(t)\dot{f}_{-}(t)-f_{-}(t)\dot{f}_{+}(t)\neq 0},
\ee
and define
\be
\ba{lll}
J_{0}:=-i\frac{1}{\,2c\,}
\left(f_{+}^{2}+f_{-}^{2}\right)\partial_{t}\,,~~~&~~~
J_{1}:=-i\frac{1}{\,2c\,}\left(f_{+}^{2}-f_{-}^{2}\right)\partial_{t}\,,~~~&~~~
J_{2}:=-i\frac{1}{\,c\,}f_{+}f_{-}\,\partial_{t}\,,
\ea
\label{Jso(1,2)}
\ee
then the  isometry of  $AdS_{2}$ or the global conformal algebra,
$\mathbf{sp}(2,\R)\equiv\so(1,2)\equiv\mathbf{sl}(2,\R)$ follows in the standard form,
\be
\ba{lll}
{}\left[J_{0},J_{1}\right]=+iJ_{2}\,,~~~~&~~~~{}\left[J_{1},J_{2}\right]=-iJ_{0}\,,~~~~&~~~~
{}\left[J_{2},J_{0}\right]=+iJ_{1}\,.
\ea
\ee
Now the above three bosonic symmetries (\ref{hidden}) can be identified as
the conformal transformations of $X$ having the conformal weight $\half$,
\be
\delta X=\delta t D_{t}X\,-\,\half\left( \partial_{t}\delta t\right)X\,+\,\phi 1\,,
\label{conformal1}
\ee
where the conformal diffieomorphism, $\delta t$, is generated by $J_{0}$, $J_{1}$, $J_{2}$
above and the inhomogenious term, $\phi$, satisfies
\be
\ddot{\phi}-\Lambda \phi+\textstyle{\frac{3}{2}}\rho\left(\partial_{t}\delta t\right)
+\dot{\rho}\,\delta t=0\,.
\label{conformal2}
\ee
Furthermore, as we show in the next section, all the  five symmetries form the $\osp(1|2,\R)$
superalgebra, where the three bosonic symmetries correspond to its even part,
$\so(1,2)$.

Apart  from the dynamical supersymmetries,  there is  the usual kinematical supersymmetry,  corresponding  to the integral constant of $\kappa(t)$,
\be
\ba{ll}
\delta \A=\delta X=0\,,~~~~~&~~~~~\delta\psi=\varepsilon\,1\,.
\ea
\label{KINE}
\ee
In parallel  to this,    there exit  two extra bosonic symmetries given by\footnote{We thank
Gordon Semenoff for pointing out the extra bosonic  symmetries.}
\be
\ba{lllll}
\delta X=f_{+}(t)1\,,~~~&~~~\delta\psi=\delta \A=0~~~&~~
\mbox{or}~~&~~~\delta X=f_{-}(t)1\,,~~~&~~~\delta\psi=\delta \A=0\,,
\ea
\ee
which can be also identified as the special case of (\ref{conformal1}), (\ref{conformal2}),
with the choice, ``\,$\delta t\equiv 0$\,".\\

Note that  the $\mathbf{su}(N)$ sector and the $\mathbf{u}(1)$ sector are completely decoupled, while   $\rho(t)$ couples to the $\mathbf{u}(1)$ sector  or the ``center of mass"  only.
~\\
~\\
~\\

\subsection{Hamiltonian and the Dirac bracket}
The Euler-Lagrangian equations read
\be
\ba{ll}
D_{t}D_{t}X-\psi\psi-{\Lambda}(t)X-\rho(t)1=0\,,~~~~~&~~~
D_{t}\psi+i\left[X,\psi\right]=0\,,\\
{}&{}\\
\left[D_{t}X,X\right]+i\psi\psi=0~~~~~&~~~:~\mbox{Gauss~constraint}\,.
\ea
\label{EOM}
\ee
Up to the Gauss constraint or the first-class constraint,  the cubic vertex term vanishes, $\tr\left(X\psi\psi\right)\simeq 0$, so that the Hamiltonian becomes simply  a harmonic oscillator type, being free of the fermion,
\be
\ba{ll}
{H}=\tr\Big[\half P^{2}-\half {\Lambda}(t) X^{2}-\rho(t)X\Big]\,,~~~~&~~~~ P:=D_{t}X\,.
\ea
\label{Hamiltonian}
\ee
In fact,  for any gauge invariant  object,
\be
\F=\tr\left[F(X,P,\psi,t)\right]\,,
\ee
the Euler-Lagrangian equations, (\ref{EOM}),  imply
\be
\ba{ll}
\dis{\frac{{\rm d} \F}{{\rm d}t}}
&=\dis{\tr\!\left[P\frac{\partial~}{\partial X }+\Big({\Lambda}(t)X+\rho(t)1\Big)\frac{\partial~}{\partial P}\right]\F-i\tr\Big(\big[X,F\big]\Big)+
\frac{\partial \F}{\partial t}}\\
{}&{}\\
{}&
\dis{=\left[\F\,,\,{H}\,\right\}_{D.B.}+ \frac{\partial\F}{\partial t }\,.}
\ea
\label{HD}
\ee
The Dirac bracket  for our matrix model is given by, after taking care of the primary second-class constraint for  the fermion~\cite{Dirac,Henneaux:1992ig},
\be
\dis{\left[\F\,,\,\G\,\right\}_{D.B.}=\frac{\partial\F~}{\partial X^{a}{}_{b}}
\frac{\partial\G~}{\partial P^{b}{}_{a}}-
\frac{\partial\F~}{\partial P^{a}{}_{b}}
\frac{\partial\G~}{\partial X^{b}{}_{a}}+i(-1)^{\#\F}
\frac{\partial\F~}{\partial \psi^{a}{}_{b}}
\frac{\partial\G~}{\partial \psi^{b}{}_{a}}\,,}
\ee
where  $a,b$ are the $N\times N$ matrix indices, while $\#\F=0$ or $1$, depending on the spin statistics of  $\F$, i.e. $0$ for the boson and $1$ for the fermion. \\

Due to the five symmetries of the action, there are five conserved quantities given by the Noether charges. For their  explicit expressions we refer   (\ref{CONSERV1}) and (\ref{CONSERV2}). \\


Since the $\mathbf{su}(N)$ and $\mathbf{u}(1)$ sectors are completely decoupled, it is convenient to introduce  the trace  over either the $\mathbf{su}(N)$ or $\mathbf{u}(1)$ sector only,
\be
\ba{l}
\trSU\Big[F(X,P,\psi)\Big]:=\tr\!\left[
F\Big(X-N^{-1}\tr(X),\,P-N^{-1}\tr(P),\,\psi-N^{-1}\tr(\psi)\Big)\right]\,,\\
{}\\
\trU1\Big[F(X,P,\psi)\Big]:=\tr\!\left[
F\Big(N^{-1}\tr(X),\,N^{-1}\tr(P),\,N^{-1}\tr(\psi)\Big)\right]\,.
\ea
\ee
Accordingly the quadratic Hamiltonian decomposes into the two distinct  pieces,
\be
{H}=\HSU+\HU1\,,
\ee
where
\be
\ba{ll}
\HSU=\trSU\Big[\half P^{2}-\half {\Lambda}(t) X^{2}\Big]\,,~~~&~~~
\HU1=\trU1\Big[\half P^{2}-\half {\Lambda}(t) X^{2}-\rho(t)X\Big]\,.
\ea
\ee
~\\
\newpage

\subsection{BPS states and  the cosmological principle}
From the supersymmetry transformations of the fermion, the BPS equations are
\be
f_{\pm}(t)D_{t}X=\dot{f}_{\pm}(t)X+\kappa_{\pm}(t) 1\,,
\label{BPSeq}
\ee
so that  the generic  BPS configurations decompose into the traceless and  $\mathbf{u}(1)$ parts,
\be
\ba{lll}
X(t)=f_{+}(t){\X}+h_{+}(t)1~~~~~&~~~\mbox{or}~~~&~~~~~
X(t)=f_{-}(t){\X}+h_{-}(t)1\,,
\label{BPSconf}
\ea
\ee
where ${\X}$ is an arbitrary  traceless constant matrix, and $h_{\pm}(t)$ are  the solutions of the first order differential equation,  $f_{\pm}\dot{h}_{\pm}=\dot{f}_{\pm}h_{\pm}+\kappa_{\pm}$,  corresponding to  the center of mass position, $N^{-1}\tr X(t)=h_{\pm}(t)$\,.\\

Since $f_{+}(t)\neq f_{-}(t)$, the BPS state preserves only one supersymmetry out of three supersymmetries (two  dynamical and one  kinematical). It is interesting to note that for  arbitrary time      dependent functions, say $f_{+}(t)$ and $h_{+}(t)$, there exits a supersymmetric matrix model where $X(t)=f_{+}(t){\X}+h_{+}(t)1$ corresponds to a  BPS state, and furthermore there exits always  its  ``twin"  BPS state given by `$\,+\rightarrow -\,\,$'.\\

Utilizing the gauge symmetry~(\ref{gaugesym}), one can diagonalize ${\X}$ in order to  show the positions of  the $N$   D-particles in the BPS sector,
\be
X(t)=\mbox{diag}\Big(x_{1}(t),\,x_{2}(t),\,\cdots,\,x_{N}(t)\Big)=f_{\pm}(t)\,
\mbox{diag}\Big({\rm x}_{1},\,{\rm x}_{2},\,\cdots,\,{\rm x}_{N}\Big)+h_{\pm}(t)1\,.
\ee
A remarkable fact is that all  D-particles have precisely  the same relative  movement, same position, same velocity, same acceleration, \textit{etc.}  up to the constant scaling  factors  which entirely depend on their initial positions or  so called  the co-moving  coordinates. This matches precisely with  the ``homothetic ansatz'' adopted in the cosmology literature in order to incorporate the \textit{cosmological principle}~\cite{Alvarez:1997fy,Freedman:2004xg}.  In fact, the second order differential equation,  $\ddot{f}_{\pm}=f_{\pm}{\Lambda}$,  (\ref{Raychaudhuri})  can be identified as the Raychaudhuri's equation in cosmology, where  ${\Lambda}$ is indeed the
time      dependent  cosmological ``constant".  Also, in the matrix approach to the cosmology~\cite{Freedman:2004xg,Gibbons:2003rv,Das:2004hw},  it is   natural to associate  $\Lambda$ as the non-relativistic cosmological constant, and associate ${\Lambda}>0$ and  ${\Lambda}<0$ with the \textit{de-Sitter} and  \textit{Anti-de-Sitter} space respectively accounting the repulsive and attractive potential.  Thus, although the geometries  of $dS_{2}$ and $AdS_{2}$ coincide, we distinguish them by the sign of $\Lambda$, throughout the paper.\\
~\\
~\\
\newpage

\section{$\osp(1|2,\R)$ superalgebra\label{SecOSP}}
After  the standard quantization,  $\left[\F\,,\,\G\,\right\}_{D.B.}\rightarrow -i\left[\F\,,\,\G\,\right\}$, the  present $\osp(1|2,\R)$ matrix model leads to  the following `\,Heisenberg $\oplus$ Clifford\,' algebra,
\be
\ba{ll}
\dis{\Big[X^{a}{}_{b}\,,\,P^{c}{}_{d}\,\Big]=i
\,\delta^{a}_{~d}\,\delta^{c}_{~b}\,,}~~~~&~~~~
\dis{\left\{\psi^{a}{}_{b}\,,\,\psi^{c}{}_{d}\,\right\}=\delta^{a}_{~d}\,\delta^{c}_{~b}\,.}
\ea
\label{QUAN}
\ee
In  Sec.~\ref{ospK},   utilizing the above  algebra alone, especially from the $\mathbf{su}(N)$ sector only,  we  construct   explicitly  the generators of the $\osp(1|2,\R)$ superalgebra.\footnote{For the construction of other various algebras, see  \cite{Gunaydin}.}  The number of odd generators is two, and this is consistent with the fact that there are two dynamical  supersymmetries  in the matrix model, rather than one.  Sec.~\ref{UIR} is devoted to the analysis on the unitary irreducible representations of the superalgebra, $\osp(1|2,\R)$.   Further analysis on the superalgebra from the dynamical   point  of view is given    in Sec.~\ref{ospD}.\\
~\\


\subsection{$\osp(1|2,\R)$ superalgebra - kinematical point of view\label{ospK}}
There are  five  \textit{real} generators in $\osp(1|2,\R)$ which we take as
\be
\ba{ll}
\QP:=\trSU\!\left(\psi P\right)\,,~~~~~&~~~~~
\QX:=\trSU\!\left(\psi X\right)\,,
\ea
\label{QPX}
\ee
and
\be
\ba{lll}
K_{0}:=\half\trSU\!\left(P^{2}+X^{2}\right)\,,~&~~~
K_{1}:=\half\trSU\!\left(P^{2}-X^{2}\right)\,,~&~~~
K_{2}:=\half\trSU\!\left(XP+PX\right)\,.
\ea
\label{K012}
\ee
Alternatively we may construct the generators out of the full $\mathbf{u}(N)$ matrices,   including the $\mathbf{u}(1)$ part and  using the ordinary trace, i.e. $\trSU\rightarrow \tr$.
All the results below will remain identical, but the resulting $\so(1,2)$ Casimir will not be a conserved time independent operator, which is not what we want. In order to account for the kinematical supersymmetry, (\ref{KINE}), we may also include one more odd generator, $Q_{\scriptstyle{\rm kinematical}}=\tr(\psi)$. However this  commutes  with any   generator above  in the $\mathbf{su}(N)$ sector.\\

All the super-commutator relations\footnote{Although  the $\osp(1,|2,\R)$ super-commutator relations  above are    direct consequences of the `\,Heisenberg $\oplus$ Clifford\,'  algebra,  the way to express  the generators  in terms of $X$, $P$ and $\psi$ is not unique.   In fact,  $\so(1,2)$ algebra was   identified  thirty years ago~\cite{deAlfaro:1976je} using a `non-polynomial'   basis in the conformal matrix model having the inverse  square potential, and based on the  observation,   Strominger  proposed  that the conformal matrix model is dual to $2D$ type 0A string theory on  $AdS_{2}$~\cite{Strominger:2003tm} (see also \cite{Ho:2004qp,Aharony:2005hm}). }
of the  $\osp(1|2,\R)$  superalgebra follow simply  from the `\,Heisenberg $\oplus$ Clifford\,'  algebra, (\ref{QUAN}),
\be
\ba{lll}
\QP^{\!2}=\half\left(K_{0}+K_{1}\right)\,,~~~~&~~~~
\QX^{\!2}=\half\left(K_{0}-K_{1}\right)\,,~~~~&~~~~
\left\{\QP,\QX\right\}=K_{2}\,,\\
{}&{}&{}\\
{}\left[K_{0},\QP\right]=+i\QX\,,~~~~&~~~~
{}\left[K_{0},\QX\right]=-i\QP\,,~~~~&~~~~
{}\left[K_{1},\QP\right]=-i\QX\,,\\
{}&{}&{}\\
{}\left[K_{1},\QX\right]=-i\QP\,,~~~~&~~~~
{}\left[K_{2},\QP\right]=+i\QP\,,~~~~&~~~~
{}\left[K_{2},\QX\right]=-i\QX\,,\\
{}&{}&{}\\
{}\left[K_{1},K_{2}\right]=-2iK_{0}\,,~~~&~~~\left[K_{0},K_{1}\right]=+2iK_{2}\,,
~~~&~~~\left[K_{2},K_{0}\right]=+2iK_{1}\,.
\ea
\label{OSP1f}
\ee
The Casimir  of the $\osp(1|2,\R)$ superalgebra reads
\be
\ba{ll}
\Cosp=\Cso+i\left[\QP,\QX\right]\,,~~~~~&~~~~~
\dis{\Big[\,\Cosp\,,\,\mbox{anything}\,\Big]=0\,,}
\ea
\label{Casimirosp}
\ee
where  the  $\so(1,2)$ Casimir  is given by
\be
\ba{ll}
\Cso&=K_{0}^{2}-K_{1}^{2}-K_{2}^{2}\\
{}&{}\\{}&=\half\left\{
\trSU\!\left(P^{2}\right),\trSU\!\left(X^{2}\right)\right\}
-\textstyle{\frac{1}{4}}\left[\trSU\!\left(XP+PX\right)\right]^{2}\\
{}&{}\\
{}&=2\left\{\QX^{\!2},\QP^{\!2}\right\}-\left\{\QX,\QP\right\}^{2}\,.
\ea
\label{Casimirso}
\ee
~\\

The \textit{root structure} of the $\osp(1|2,\R)$ superalgebra can be identified by complexifying the generators as\footnote{For  further analysis by us   on the root structures of  super Lie algebras, see  \cite{Kim:2002zg,Lee:2004jx}.}
\be
\ba{ll}
Q_{+}:=\QP+i\QX\,,    ~~~~~&~~~~~Q_{-}:=\QP-i\QX=Q_{+}^{\dagger}\,,\\
{}&{}\\
K_{+}:=K_{1}+iK_{2}\,,~~~~~&~~~~~K_{-}:=K_{1}-iK_{2}=K_{+}^{\,\dagger}\,.
\ea
\ee
The Cartan subalgebra has only one element, $K_{0}$, and all others are either  raising,  $Q_{+},K_{+}$, or lowering,  $Q_{-},K_{-}$, operators to satisfy
\be
\ba{lll}
Q_{+}^{\,2}=K_{+}\,,~~~~&~~~~Q_{-}^{\,2}=K_{-}\,,~~~~&~~~~
\left\{Q_{-},Q_{+}\right\}=2K_{0}\,,\\
{}&{}&{}\\
{}\left[K_{0},Q_{+}\right]=+Q_{+}\,,~~~~&~~~~
{}\left[K_{+},Q_{+}\right]=0\,,~~~~&~~~~{}\left[K_{-},Q_{+}\right]=+2Q_{-}\,,\\
{}&{}&{}\\
{}\left[K_{0},Q_{-}\right]=-Q_{-}\,,~~~~&~~~~
{}\left[K_{+},Q_{-}\right]=-2Q_{+}\,,~~~~&~~~~{}\left[K_{-},Q_{-}\right]=0\,,\\
{}&{}&{}\\
{}\left[K_{0},K_{+}\right]=+2K_{+}\,,~~~~&~~~~{}\left[K_{0},K_{-}\right]=-2K_{-}\,,~~~~&~~~~
{}\left[K_{-},K_{+}\right]=4K_{0}\,.
\ea
\ee
In the Cartan  basis, the Casimir operators, (\ref{Casimirosp}), (\ref{Casimirso}),  read
\be
\ba{ll}
\Cosp=\Cso+\half\left[Q_{-},Q_{+}\right]\,,~~~~~&~~~~
\Cso=K_{0}^{2}-\half\left\{K_{-},K_{+}\right\}\,.
\ea
\label{Cso+-}
\ee
~\\

The $\osp(1|2,\R)$   superalgebra can be represented by $(2+1)\times(2+1)$ real supermatrices, $M$, satisfying
\be
\ba{ll}
M^{T}{\cal J}+{\cal J}M=0\,,~~~~&~~~~\dis{{\cal J}=\left(\ba{rrr}0&~1~&0\\-1&~0~&0\\0&~0~&1\ea
\right)\,,}
\ea
\ee
so that its generic form reads, with the  even and odd real Grassmann entries, $x$, $\theta$, \be
\dis{M=\left(\ba{lll}
~x_{2}&-x_{+}~&~\theta_{1}\\
~x_{-}&-x_{2}~&~\theta_{2}\\
-\theta_{2}~&~~\theta_{1}~&~0~\ea
\right)\,.}
\ee
Note that the  $2\times 2$ bosonic part corresponds to  $\mathbf{sp}(2,\R)\equiv\so(1,2)\equiv\mathbf{sl}(2,\R)$, as it corresponds to $x_{\mu}\gamma^{\mu}$, where  $~x_{\pm}=x_{0}\pm x_{1}$,  and $\gamma^{\mu}$ is the  $\so(1,2)$ gamma matrix,
\be
\ba{lll}
\gamma^{0}:=\left(\ba{rr}0&-1\\1&0\,\ea\right),~~&~~
\gamma^{1}:=\left(\ba{rr}0\,&-1\\-1&0\,\ea\right),~~&~~
\gamma^{2}:=\left(\ba{rr}1&0\,\\0&-1\ea\right),\\
{}&{}&{}\\
\multicolumn{3}{c}
{\gamma^{\mu}\gamma^{\nu}+\gamma^{\nu}\gamma^{\mu}=2\eta^{\mu\nu}\,,
~~~~~~~~~\eta=\diag(-++)\,.}
\ea
\label{so12gm}
\ee

In fact, with the notion  of a $(1+2)$-dimensional  two component Majorana  spinor and its `charge conjugate',
\be
\ba{ll}
Q=\left(\ba{l}Q_{1}\\ Q_{2}\ea\right)
=\left(\ba{l}\QP\\ \QX\ea\right)\,,~~~~&~~~~
\bar{Q}:=Q^{T}\gamma^{0}=\left(~\QX~~-\QP~\right)\,,
\ea
\ee
the $\osp(1|2,\R)$ superalgebra, (\ref{OSP1f}),  can be rewritten in a compact form,
\be
\ba{lll}
\left\{Q\,,\,\bar{Q}\,\right\}=\gamma^{\mu}K_{\mu}\,,~~~~&~~~~
\left[\,K_{\mu}\,,\,Q\,\right]=i\gamma_{\mu}Q\,,~~~~&~~~~
\left[\,K_{\mu}\,,\,K_{\nu}\,\right]=2i\,\epsilon_{\mu\nu\lambda}K^{\lambda}\,,
\ea
\label{COMPACT}
\ee
where $\epsilon_{\mu\nu\lambda}$ is the usual  three form with $\epsilon_{012}\equiv 1$.    Note also, from $\bar{Q}Q=\left[\QX,\QP\right]$, that  the $\osp(1|2,\R)$ Casimir operator,  $\Cosp$, (\ref{Casimirosp}) is indeed manifestly $\mathbf{SL}(2,\R)$ invariant. \\

In a similar fashion to  above, one can also equip the bosonic operators  with the    $\mathbf{SL}(2,\R)$ covariant structure,
\be
\ba{ll}
{\cal V}=\left(\ba{l}{\cal V}_{1}\\ {\cal V}_{2}\ea\right):=\left(\ba{l}P\\X\ea\right),~~~~&~~~~
K_{\mu}=\half\bar{{\cal V}}\gamma_{\mu}{\cal V}\,.
\ea
\ee
~\\

\subsection{Unitary irreducible representations of $\osp(1|2,\R)$ \label{UIR}}
In order to  analyze  the {\textit{unitary irreducible representations}}  or {\textit{unitary supermultiplets}}    of the {$\osp(1|2,\R)$ superalgebra spanned by the five \textit{real} generators,   (\ref{QPX}), (\ref{K012}), one needs to take    $K_{0}$  as the `good' quantum number operator to  diagonalize  it. Different  choice of the good quantum number operator, e.g.   $K_{2}$, is not  compatible with the unitarity, as it would lead to the raising and lowering operators with the pure imaginary unit, such as $[K_{2}, (K_{1}\pm K_{0})]=\pm 2i(K_{1}\pm K_{0})$.\\

 Any  $\osp(1|2,\R)$ supermultiplet decomposes into  $\so(1,2)$ multiplets.  We first review briefly  the general properties of the latter or the unitary irreducible representations    of  $\so(1,2)$.\footnote{For further analysis see  e.g. \cite{so(12)}.}
From (\ref{Cso+-}) and  the commutator relations, we get
\be
\dis{\Cso+1+K_{\pm}K_{\mp}=(K_{0}\mp1)^{2}\,.}
\label{soui}
\ee
From  the Hermitian conjugacy  property, $K_{+}=K_{-}^{\dagger}$,  the third term on the left hand side, $K_{\pm}K_{\mp}$,  is positive semi-definite, while  the possible  minimum value of the right hand side for the states in a unitary  irreducible representation may lie
\be
0\leq\mbox{min}\!\left[(K_{0}\mp1)^{2}\right]\leq 1\,,
\ee
\textit{if} the raising or lowering operators  act nontrivially ever. But, this is impossible when $\Cso>0$. In this case, the unitary representation is infinite dimensional and  characterized by the existence of either the lowest weight state obeying
\be
\ba{llll}
K_{-}|l,l\rangle=0\,,~~~&~~~~K_{0}|l,l\rangle=l|l,l\rangle\,,~~~&~~~
\Cso=l(l-2)>0\,,~~~&~~l>2\,,
\ea
\ee
or the highest weight state obeying
\be
\ba{llll}
K_{+}|h,h\rangle=0\,,~~~&~~~~K_{0}|h,h\rangle=h|h,h\rangle\,,~~~&~~~
\Cso=h(h+2)>0\,,~~~&~~h<-2\,.
\ea
\ee
When $\Cso=0$,  there exists only one trivial state, $|0,0\rangle$, satisfying
\be
\ba{ll}
K_{\pm}|0,0\rangle=0\,,~~~~&~~~~K_{0}|0,0\rangle=0\,.
\ea
\ee
When $-1\leq\Cso<0$, the representation is called the `continuous principal series'. It is infinite dimensional, and the lowest or highest weight state may or may not exit.
If there is a lowest or highest  weight state, then its  good quantum number   is  $+1\pm\sqrt{\Cso+1}$ or $-1\pm\sqrt{\Cso+1}$, respectively but not simultaneously. When $\Cso<-1$, there must be neither lowest  nor highest weight state, and the representation is called  the `continuous supplementary series'.\\

As for the present $\osp(1|2,\R)$ matrix model, $K_{0}$ is positive definite for the unitary multiplets  as
\be
\ba{ll}
\multicolumn{2}{l}{K_{0}=\half\trSU\!\left(P^{2}+X^{2}\right)=
\trSU\left(A^{\dagger}A\right)+\half\left(N^{2}-1\right)\,\geq\,\half \left(N^{2}-1\right)\,,}\\
{}&{}\\
\dis{{A}:=\textstyle{\frac{1}{\sqrt{2}}}\left(P- iX\right)\,,}~~~~&~~~~
\dis{\Big[A^{a}{}_{b}\,,\,A^{\dagger}{}^{c}{}_{d}\,\Big]=
\delta^{a}_{~d}\,\delta^{c}_{~b}\,.}
\ea
\label{K0positive}
\ee
Thus there exits always a lowest weight state in any $\so(1,2)$  multiplet, and from (\ref{soui}),  the $\so(1,2)$ Casimir is bounded below\footnote{In fact, from (\ref{Casimirso}),  expressing  $\Cso$  in terms of the odd generator,    the  trace of $\Cso$  also `formally' shows the positiveness,
\[
\dis{\Tr\,\Cso\approx\Tr\left(-\left[\QX,\QP\right]^{2}\right)\,\geq \,0\,.}
\]
The subtlety is due to the infinite sum over the infinite dimensional $\so(1,2)$ multiplet. }
\be
\ba{ll}
\Cso\geq\textstyle{\frac{1}{4}}(N^{2}-1)(N^{2}-5)~~~~~&~\mbox{for~~}N\geq 3\,,\\
{}&{}\\
\Cso>0~~~~\mbox{or}~~~~\Cso=-\textstyle{\frac{3}{4}}~~~~&~\mbox{for~~}N=2\,.
\ea
\label{POSITIVE}
\ee
~\\

Now as for the $\osp(1|2,\R)$ unitary supermultiplet, we first note that the odd roots, $Q_{\pm}$,  shift  the `good' quantum number by one unit, half of what $K_{\pm}$ do. Hence the odd roots move one $\so(1,2)$ multiplet to another  inside a   $\osp(1|2,\R)$ supermultiplet, but at most once due to  $Q_{\pm}^{2}=K_{\pm}$. Similar to (\ref{soui}), we also have
\be
\dis{\Cosp+K_{\pm}K_{\mp}\pm Q_{\pm}Q_{\mp}=K_{0}\left(K_{0}\mp 1\right)\,.}
\ee
After all, utilizing all the facts above,  we conclude  that any unitary irreducible   representation of the  $\osp(1|2,\R)$ superalgebra satisfying the positiveness, (\ref{K0positive}), is  infinite dimensional and  characterized by the existence of the super-lowest weight state obeying
\be
\ba{llll}
Q_{-}|l_{s},l_{s}\rangle=0\,,~~~&~~~K_{0}|l_{s},l_{s}\rangle=l_{s}|l_{s},l_{s}\rangle\,,
~~~&~~
\Cosp=l_{s}(l_{s}-1)\,,~~&~~l_{s}\geq \half \left(N^{2}-1\right)\,.
\ea
\ee
Furthermore, the $\osp(1|2,\R)$ unitary supermultiplet always decomposes into two $\so(1,2)$ multiplets whose lowest weight states are given by
\be
\ba{lll}
|l_{s},l_{s}\rangle~~~~~~~&\mbox{and}&~~~~~~~
\dis{|{l_{s}+1},{l_{s}+1}\rangle=\frac{1}{\sqrt{2l_{s}}}\,Q_{+}|l_{s},l_{s}\rangle\,.}
\ea
\ee
~\\

\subsection{$\osp(1|2,\R)$ superalgebra - dynamical  point of view\label{ospD}}
The Noether charges corresponding to the two dynamical supersymmetries, (\ref{SUSYf2}), decompose into the $\mathbf{su}(N)$ and $\mathbf{u}(1)$ parts,
\be
\ba{l}
\tr\!\left(i\psi\delta_{\pm}\psi\right)=if_{\pm}(t)\Big(
\Q_{{\scriptscriptstyle{\rm su}(N)}}^{\pm}+\Q_{{\scriptscriptstyle{\rm u}(1)}}^{\pm}\Big)\varepsilon_{\pm}\,,\\
{}\\
\Q_{{\scriptscriptstyle{\rm su}(N)}}^{\pm}:=\trSU\!\left[\psi\Big(P-g_{\pm}(t)X\Big)\right]\,,\\
{}\\
\Q_{{\scriptscriptstyle{\rm u}(1)}}^{\pm}:=\trU1\!
\left[\psi\Big(P-g_{\pm}X-\left(\kappa_{\pm}/f_{\pm}\right)1\Big)\right]\,.
\ea
\label{Noethersuper}
\ee
where we put
\be
\ba{ll}
g_{\pm}(t):=\dis{\frac{\dot{f}_{\pm}(t)}{f_{\pm}(t)}}\,,
~~~~&~~~~\dot{g}_{\pm}+g_{\pm}^{2}={\Lambda}(t)\,.
\ea
\label{gpmL}
\ee
Because the Hamiltonian as well as the above two supercharges in the $\mathbf{su}(N)$ sector  can be expressed in terms of    the previous ``kinematical" basis, $\QX,\QP,K_{0},K_{1},K_{2}$, (\ref{QPX}), (\ref{K012}),   the underlying supersymmetry algebra must correspond to $\osp(1|2,\R)$, no matter what the dynamics is. However,  the use of the above  supercharges, $\Q_{{\scriptscriptstyle{\rm su}(N)}\pm}$, will  not lead to   simple expressions for  the superalgebra. For example, from the conservation of the Noether charge and  Eq.(\ref{HD}), the commutator relation between the Hamiltonian and the supercharge reads in a less economic manner,
\be
\dis{\left[H\,,\,\Q_{{\scriptscriptstyle{\rm su}(N)}}^{\pm}\right]=
ig_{\pm}\Q_{{\scriptscriptstyle{\rm su}(N)}}^{\pm}
+i\,\frac{\dot{g}_{\pm}}{\,\,g_{+}-g_{-}}
\left(\Q_{{\scriptscriptstyle{\rm su}(N)}}^{+}-
\Q_{{\scriptscriptstyle{\rm su}(N)}}^{-}\right) \,.}
\ee
~\\

Henceforth, in order to analyze  the underlying $\osp(1|2,\R)$ superalgebra in a simple fashion  but still to keep track of the dynamical properties, we slightly modify the basis  of  the odd generators and  keep the Hamiltonian explicitly as a $\so(1,2)$ generator. Note that the change of basis requires the time      dependent coefficients due  to  $\Lambda(t)$, as $\Q_{{\scriptscriptstyle{\rm su}(N)}}^{\pm}=\QP-g_{\pm}(t)\QX$. Hence, only  with specific   time      dependent coefficients we can write down the  time      independent  conserved quantities, as one can expect from (\ref{HD}).
All together there are  five  conserved  ``true" Noether charges corresponding to  the five symmetries, (\ref{SUSYf2}), (\ref{hidden}). Namely we have   the two fermionic  conserved Noether charges for the two dynamical supersymmetries,
\be
f_{\pm}\Q_{{\scriptscriptstyle{\rm su}(N)}}^{\pm}= f_{\pm}\QP-\dot{f}_{\pm}\QX\,,
\label{CONSERV1}
\ee
and   three bosonic conserved Noether charges for the $\so(1,2)$ symmetries  (\ref{hidden}),
\be
\ba{l}
\left(f_{\pm}\Q_{{\scriptscriptstyle{\rm su}(N)}}^{\pm}\right)^{2}=\half\left(f_{\pm}^{2}+\dot{f}_{\pm}^{2}\right)K_{0}+
\half\left(f_{\pm}^{2}-\dot{f}_{\pm}^{2}\right)K_{1}-f_{\pm}\dot{f}_{\pm}K_{2}\,,\\
{}\\
\left\{f_{+}\Q_{{\scriptscriptstyle{\rm su}(N)}}^{+}
\,,\,f_{-}\Q_{{\scriptscriptstyle{\rm su}(N)}}^{-}\right\}=
\left(f_{+}f_{-}+\dot{f}_{+}\dot{f}_{-}\right)K_{0}+
\left(f_{+}f_{-}-\dot{f}_{+}\dot{f}_{-}\right)K_{1}-
\left(f_{+}\dot{f}_{-}+f_{-}\dot{f}_{+}\right)K_{2}.
\ea
\label{CONSERV2}
\ee
Apart from the above five Noether charges, both of the  $\osp(1|2,\R)$ and $\so(1,2)$ Casimir operators, $\Cosp$ (\ref{Casimirosp}) and  $\Cso$ (\ref{Casimirso}), are   also conserved time independent quantities,  since they do not include any explicit time dependency and they  commute with the Hamiltonian, for sure. \\
~\\

\subsubsection{ $\osp(1|2,\R)$ superalgebra when $\Lambda(t)\neq 0$}
In a similar fashion to  the standard harmonic oscillator analysis, we first set a pair of   operators,\footnote{In fact, when $\Lambda(t)$ is constant,  $A_{\pm}$ correspond to the  generators of $W_{\infty}$  algebra~\cite{Winfinity}.}
\be
\dis{A_{\pm}(t):=\frac{\,P\pm \sqrt{{\Lambda}(t)}\,X\,}{\sqrt{2}}\,,}
\ee
and define a pair of    even  generators in $\osp(1|2,\R)$ by
\be
\dis{J_{\pm}(t):=\trSU\Big[A_{\pm}(t)^{2}\Big]\,,}
\ee
as well as  a pair of    odd   generators,
\be
\dis{\Q_{\pm}(t):=\trSU\Big[\psi A_{\pm}(t)\Big]\,.}
\ee
Note that ${\Q}_{\pm}$ coincide with the actual supercharges, $\Q_{{\scriptscriptstyle{\rm su}(N)}}^{\pm}$,  (\ref{Noethersuper}),   provided that  ${\Lambda}(t)$ is  constant.\\

The Hamiltonian for the $\mathbf{su}(N)$ sector is then
\be
\dis{\HSU=\half\trSU\Big[A_{+}(t)A_{-}(t)+A_{-}(t)A_{+}(t)\Big]=
\trSU\Big[A_{\pm}(t)A_{\mp}(t)\Big]\mp\half i\sqrt{\Lambda(t)}\left(N^{2}-1\right),}
\label{HSUL}
\ee
and from the quantization relation,
\be
\dis{\Big[A_{-}(t)^{a}{}_{b}\,,\,A_{+}(t)^{c}{}_{d}\,\Big]=-i\sqrt{\Lambda(t)}\,
\delta^{a}_{~d}\,\delta^{c}_{~b}\,,}
\ee
we obtain such as
\be
\ba{lll}
\dis{\left[{\HSU},\hat{A}_{\pm}\right]=\mp i\sqrt{\Lambda(t)}\,\hat{A}_{\pm}}\,,~&~
\dis{\left[J_{-},\hat{A}_{+}\right]=-2i\sqrt{\Lambda(t)}\,\hat{A}_{-}\,,}~&~
\dis{\left[J_{+},\hat{A}_{-}\right]=+2i\sqrt{\Lambda(t)}\,\hat{A}_{+}\,,}
\ea
\label{HApm}
\ee
where we set
\be
\hat{A}_{\pm}:=A_{\pm}-N^{-1}\tr\left(A_{\pm}\right)1\,.
\ee

Now, the $\osp(1|2,\R)$ superalgebra reads in terms of   ${\Q}_{\pm},J_{\pm},{\HSU}$,

\be
\ba{lll}
{\Q}_{+}^{\,2}=\half J_{+}\,,&
{\Q}_{-}^{\,2}=\half J_{-}\,,&
{}\left\{{\Q}_{+},{\Q}_{-}\right\}={\HSU}\,,\\
{}&{}&{}\\
{}\left[J_{-},{\Q}_{+}\right]=-2i\sqrt{\Lambda(t)}\,{\Q}_{-}\,,&
{}\left[J_{+},{\Q}_{-}\right]=+2\sqrt{\Lambda(t)}\,{\Q}_{+}\,,&
{}\left[{\HSU},{\Q}_{\pm}\right]=\mp i\sqrt{\Lambda(t)}\, {\Q}_{\pm}\,,\\
{}&{}&{}\\
{}\left[J_{-},J_{+}\right]=-4i\sqrt{\Lambda(t)}\,{\HSU}\,,&\,
\left[{\HSU},J_{\pm}\right]=\mp 2i\sqrt{\Lambda(t)}\,J_{\pm}\,,&\,
\left[J_{\pm},{\Q}_{\pm}\right]=0\,.
\ea
\label{superalgebranot0}
\ee
Especially, the $\so(1,2)$ Casimir operator, (\ref{Casimirso}), can be reexpressed as
\be
\dis{\Cso=K_{0}^{2}-K_{1}^{2}-K_{2}^{2}=\frac{1}{\Lambda(t)}\Big[
\half\left\{J_{+},J_{-}\right\}-{\HSU}^{2}\Big]\,.}
\label{CasimirJJH}
\ee
From $J_{\pm}^{\dagger}=J_{\pm}$ for $\Lambda>0$ and  $J_{+}^{\dagger}=J_{-}$ for $\Lambda<0$, there exits a $\mathbf{SO}(1,2)$ rotation which transforms  $\HSU$ to  $K_{1}$ if $\Lambda>0$ or $K_{0}$ if $\Lambda<0$.\\

\subsubsection{ $\osp(1|2,\R)$ superalgebra when $\Lambda(t)=0$}
 If ${\Lambda}=0$, $A_{+}$ coincides with $A_{-}$, and the above super-commutator relations (\ref{superalgebranot0}) do not faithfully represent the super Lie algebra, $\osp(1|2,\R)$. In order to do so, one needs to define the generators differently. When $\Lambda=0$ we have
$f_{+}=1$, $f_{-}=t$, and   the corresponding two supercharges, (\ref{Noethersuper}), in the $\mathbf{su}(N)$ sector  are
\be
\ba{ll}
\Q_{{\scriptscriptstyle{{\Lambda}=0}}}^{+}=\trSU\!\left[\psi P\right]\,,~~~~&~~~~\Q_{{\scriptscriptstyle{{\Lambda}=0}}}^{-}
=\trSU\!\left[\psi(P-t^{-1}X)\right]\,,
\ea
\ee
while the Hamiltonian is given by
\be
{\HSU}=\half\trSU\!\left(P^{2}\right)=\half\left(K_{0}+K_{1}\right)\,.
\ee
Rather than $\Q_{{\scriptscriptstyle{{\Lambda}=0}}}^{\pm}$, we adopt the kinematical odd generators, (\ref{QPX}),
\be
\ba{ll}
\QP=\trSU\!\left(\psi P\right)=\Q_{{\scriptscriptstyle{{\Lambda}=0}}}^{+}\,,~~~~&~~~~
\QX=\trSU\!\left(\psi X\right)=t\left(\Q_{{\scriptscriptstyle{{\Lambda}=0}}}^{+}
-\Q_{{\scriptscriptstyle{{\Lambda}=0}}}^{-}\right)\,,
\ea
\ee
and write the $\osp(1|2,\R)$ superalgebra in terms of the real basis,
\be
\ba{lll}
\QP^{\!2}=\HSU\,,~~&~~\QX^{\!2}={\VSU}:=\half\trSU\!\left(X^{2}\right)
\,,~~&~~\left\{\QP,\QX\right\}=K_{2}\,,\\
{}&{}&{}\\
{}\left[\HSU,\QX\right]=-i\QP\,,~~&~~{}\left[\HSU,\QP\right]=0\,,~~&~~
{}\left[{\VSU},\QX\right]=0\,,\\
{}&{}&{}\\
{}\left[{\VSU},\QP\right]=+i\QX\,,~~&~~
{}\left[K_{2},\QX\right]=-i\QX\,,~~&~~
{}\left[K_{2},\QP\right]=+i\QP\,,\\
{}&{}&{}\\
{}\left[\HSU,{\VSU}\right]=-iK_{2}\,,~~&~~\left[K_{2},\HSU\right]=+2i\HSU\,,
~~&~~\left[K_{2},{\VSU}\right]=-2i{\VSU}\,.
\ea
\ee
In particular, the $\so(1,2)$ Casimir operator, (\ref{Casimirso}), reads
\be
\dis{\Cso}=\dis{K_{0}^{2}-K_{1}^{2}-K_{2}^{2}
=2\left\{\HSU\,,\, {\VSU}\right\}-K_{2}^{2}}\,.
\label{CsoT0}
\ee
~\\

\section{Discussion and conclusion\label{CONCLUSION}}
We have derived a $\N=2$ supersymmetric matrix model, (\ref{FINAL2}), with  quadratic and linear  potentials whose coefficients are  arbitrary time      dependent   `cosmological constant',  $\Lambda(t)$,  and `electric flux background', $\rho(t)$. The  matrix model corresponds to the most general supersymmetric deformations of the   matrix quantum mechanics having the $2D$  super Yang-Mills origin.  We have shown that, for arbitrary $\Lambda(t)$ and $\rho(t)$,   the matrix model enjoys  two dynamical supersymmetries, $Q_{1}$, $Q_{2}$,  and three  bosonic symmetries, $K_{0},K_{1},K_{2}$,  which amount to the superalgebra, $\osp(1|2,\R)$, (\ref{COMPACT}),
\be
\ba{lll}
\left\{Q\,,\,\bar{Q}\,\right\}=\gamma^{\mu}K_{\mu}\,,~~~~&~~~~
\left[\,K_{\mu}\,,\,Q\,\right]=i\gamma_{\mu}Q\,,~~~~&~~~~
\left[\,K_{\mu}\,,\,K_{\nu}\,\right]=2i\,\epsilon_{\mu\nu\lambda}K^{\lambda}\,.
\ea
\label{superalgebraCON}
\ee
If  the matrix model had only one supersymmetry as in the $2D$ minimal super Yang-Mills, the $\osp(1|2,\R)$ structure would be absent.\\

The  matrix model is to describe   the noncritical $3D$ $\M$-theory  on generic supersymmetric  backgrounds in a controllable manner through the matrix regularization, and our claim is further that,  with the arbitrariness of $\Lambda(t)$ and $\rho(t)$,  it  also provides   holographic duals to   various two dimensional  superstring theories, as we argue below.\\
\\

\subsection{Normalizable and non-normalizable wave functions \label{QuantumWaveFunction}}
At the quantum level, the wave function  satisfies
\be
\dis{\tr\Big[\half P^{2}-\half {\Lambda}(t) X^{2}-\rho(t)X\Big]\left|\Psi(t)\rangle\right.=
i\frac{\partial~}{\partial t}\,\left|\Psi(t)\rangle\right.\,.}
\ee
As is well known, when $\Lambda(t)$ is negative, the potential is bounded below and the \textit{normalizable} or \textit{unitary} wave functions have the discrete  spectrum. Namely, from (\ref{HApm}), the lowering operator, $\hat{A}_{-}$ lowers  the eigenvalue of $\HSU$ by the unit $\sqrt{\left|\Lambda\right|}$. However, since  $\HSU$ should be positive definite for the unitary representations, there must be a ground state which is annihilated by $\hat{A}_{-}$,
\be
\ba{ll}
\hat{A}_{-}\left|0\rangle\right.=0\,,~~~~&~~~~
\dis{\left|0\rangle\right.=e^{-\half{\sqrt{\left|\Lambda\right|}}{}\,\tr X^{2}}\left|P\equiv 0\rangle\right.\,.}
\ea
\ee
All other excited states are then constructed by acting the raising operator, $\hat{A}_{+}$ to the ground state.   Due to the Gauss constraint,  one  needs to restrict  on the gauge singlets, which can be simply done by   taking the  `trace' of the $\mathbf{u}(N)$ indices in all the possible  ways~\cite{Park:2001sj}.  The quantum states in the Hilbert space then form  the unitary irreducible representations  of the  superalgebra, $\osp(1|2,\R)$, and in particular, their $\so(1,2)$ Casimir is positive definite for $N\geq 3$,  (\ref{POSITIVE}).  The energy  spectrum is discretized by the unit $\sqrt{\left|\Lambda(t)\right|}$,  and the  zero point vacuum energy is,   from (\ref{HSUL}),    $\half \sqrt{\left|\Lambda(t)\right|}\left(N^{2}-1\right)$.  The vacuum has the degeneracy, $2^{[N^{2}/2]}$, due to the fermions.  The non-vanishing zero point energy refers  to the existing  two other bosonic charges in the superalgebra apart from the Hamiltonian.  \\

On the other hand, when $\Lambda(t)$ is positive, the wave functions can not be normalizable, as    the  raising and lowering operators  shift the energy spectrum  by  the imaginary  unit, $i\sqrt{\Lambda}$, while $\HSU$ should have real eigenvalues for the normalizable states.  Physically, this  amounts to the fact that the matrix model describes the Fermi sea  (see e.g.  \cite{Karczmarek:2003pv,Das:2004hw}). \\

Therefore, in order to have a unifying description for arbitrary  $\Lambda(t)$,  the full Hilbert space of the $\M$-theory   matrix model should include not only     the \textit{normalizable states}  but also   the \textit{non-normalizable states}, allowing both the unitary and the non-unitary representations of $\osp(1|2,\R)$.  The former is relevant only to the case  $\Lambda(t)<0$.    Without the   concern    about   the normalizability,  the following Schr\"{o}dinger  equation  has always solutions for arbitrary  energy, $E(t)$,
\be
\dis{-\tr\left[\half\!\left( \frac{\partial~}{\partial X}\right)^{2}+\half {\Lambda}(t) X^{2}+\rho(t)X\right]\Psi(X,t)=E(t)\Psi(X,t)\,.}
\ee
In particular, the momentum operator $P=-i\frac{\partial~}{\partial X}$ is no longer  necessarily \textit{real}. Again the rasing and lowering operators, $\hat{A}_{\pm}$, generate new solutions with the  shifted  energy, $E(t)\mp i\sqrt{\Lambda(t)}$. \\

The reason to consider the phase space over the complex planes rather than the real lines is manifest in the path integral formalism, since when $\Lambda(t)>0$, the local minima of the Hamiltonian are located  on the genuine complex planes rather than the real lines so that one should take \textit{$P$  Hermitian  and $X$  anti-Hermitian, or vice versa.}\\

All the BPS states are non-normalizable or non-unitary :    From its defining property,
\be
\ba{ll}
Q_{{\scriptscriptstyle \rm BPS }}\left|{\rm BPS}\rangle\right.=0\,,~~~~&~~~~
Q_{{\scriptscriptstyle \rm BPS }}=\tr\!\left[\psi\left(f P-\dot{f}X-\kappa 1\right)\right]
=Q_{{\scriptscriptstyle \rm BPS }}^{\dagger}\,,
\ea
\ee
and the positive definite property of $\textstyle{Q_{{\scriptscriptstyle \rm BPS }}^{2}=
\half\tr\!\left(f P-\dot{f}X-\kappa 1\right)^{\!2}}$ for the unitary states,
if the BPS state were normalizable,
it would mean $f P\left|{\rm BPS}\rangle\right.=
\left(\dot{f}X+\kappa 1\right)\left|{\rm BPS}\rangle\right.$ so that
${\left.\langle{\rm BPS}\right|fXP\left|{\rm BPS}\rangle\right.=
\left.\langle{\rm BPS}\right|\left(\dot{f}X^{2}+\kappa X\right)
\left|{\rm BPS}\rangle\right.=\left.\langle{\rm BPS}\right|fPX\left|{\rm BPS}\rangle\right.}$.
But  this clearly contradicts with the quantization, $[X,P]\neq 0$. \\

\subsection{Projection to the `$\so(1,2)$ hyperspace'}
We consider a  projection map from   the full  phase space  to the   $`\so(1,2)$ hyperspace' given by the  ``coordinates", $K_{0}$, $K_{1}$, $K_{2}$, (\ref{K012}). The induced dynamics therein is subject to
\be
\ba{ll}
\dis{\dot{K}_{\mu}=i\left[\HSU\,,\,K_{\mu}\right]=2\epsilon_{\mu\nu\lambda}
{\cal T}^{\nu}K^{\lambda}\,,}~~~&~~~
{\cal T}^{\nu}:=\left(\,\half(1-\Lambda)\,,\,\half(1+\Lambda)\,,\,0\,\right)\,,\\
{}&{}\\
{\cal T}^{\mu}K_{\mu}=\HSU\,,~~~&~~~{\cal T}^{\mu}{\cal T}_{\mu}=\Lambda\,.
\ea
\label{Kdynamics}
\ee
Naturally, as seen from  (\ref{so12gm}), the $\so(1,2)$ hyperspace is equipped with the $\so(1,2)$ metric, $\eta=\diag(-++)$.  The $\so(1,2)$ Casimir, $\Cso$, (\ref{Casimirso}) highlights   the geometrical picture,
\be
\Cso=K_{0}^{2}-K_{1}^{2}-K_{2}^{2}\,.
\ee
$\Cso$ is   a conserved time independent operator,  since it does not include any explicit time dependency and it  commutes with the Hamiltonian, just like the $\osp(1|2,\R)$ Casimir.  Classically, this can be also  seen,  from (\ref{Kdynamics}),  as  $K^{\mu}\dot{K}_{\mu}=0$.   Therefore,  we observe that
\textit{for each $\so(1,2)$ multiplet in the Hilbert space,   the corresponding  $\so(1,2)$ hyperspace dynamics is constrained  on  a two dimensional  rigid surface}    such that
\be
\ba{ll}
\mbox{Euclidean\,~} dS_{2}/AdS_{2}~~~~&~~~~\mbox{if~~~~~} \Cso>0\,,\\
{}&{}\\
\mbox{Minkowskian\,~}dS_{2}/AdS_{2}~~~~&~~~~\mbox{if~~~~~} \Cso<0\,,\\
{}&{}\\
{{Null~cone}}~~~~&~~~~\mbox{if~~~~~}\Cso=0\,.
\ea
\ee
Surely the specific value of the  Casimir for each multiplet is to be superselected just like any boundary condition in quantum field theories. This also fits into the $3D$ $\M$-theory picture, to include or provide holographic dual descriptions to  all the superstring theories. The richness of the $\osp(1|2,\R)$ $\M$-theory  matrix model originates from the arbitrariness of the cosmological constant, ${\Lambda}(t)$ and  the electric  flux background, $\rho(t)$ as well as the vast amount of existing  $\so(1,2)$ multiplets  in the Hilbert  space each of which has its own two dimensional geometry. \\

However, if we restrict on the unitary irreducible representations, i.e.
the normalizable sector relevant to the case   $\Lambda(t)<0$,   we have the
 bound for the Casimir, (\ref{POSITIVE}),
\be
\ba{ll}
\Cso\geq\textstyle{\frac{1}{4}}\left(N^{2}-1\right)\left(N^{2}-5\right)\,.
\ea
\label{EVID1}
\ee
Thus, the corresponding  geometry is always Euclidean $dS_{2}/AdS_{2}$ if $N\geq 3$.
 As for the non-normalizable or  non-unitary  sector,   the above bound does not hold.\\

The bound can be also understood classically as
\be
\ba{ll}
\Cso&=\dis{\trSU\!\left(P^{2}\right)\trSU\left(X^{2}\right)-\Big[\trSU\!
\left(PX\right)\Big]^{2}}\\
{}&{}\\
{}&\dis{=
\trSU\left(P^{2}\right)\trSU\!
\left[X-\frac{\trSU\left(PX\right)}{\trSU\left(P^{2}\right)}\,P\right]^{2}\,.}
\ea
\ee
This is \textit{positive semi-definite if both $X$ and $P$ are Hermitian},
as is the case for  the expectation values of the unitary states.
 Otherwise, of course,  not.  Especially, \textit{when $P$ is Hermitian
  and $X$ is anti-Hermitian or vice versa, as in the path integral formalism
  for  $\Lambda>0$, $\Cso$ is negative semi-definite},
  implying  the Minkowskian $dS_{2}/AdS_{2}$ geometry.
  From  (\ref{gpmL}),  among the on-shell  configurations,  only the
  BPS configurations saturate the bound, $\Cso=0$. \\

From  (\ref{CasimirJJH}), (\ref{Kdynamics}),  a `dispersion relation' follows
\be
\dis{\dot{K}_{\mu}\dot{K}^{\mu}=4\Big(\HSU^{2}-\Lambda K_{\mu}K^{\mu}\Big)=
4\Big(\HSU^{2}+\Lambda(t)\Cso\Big)=2\Big\{J_{+}(t)\,,\,J_{-}(t)\Big\}\,,}
\ee
which shows that the  `mass' is conserved  if  $\Lambda(t)$ is constant.
Furthermore,  we have the positive semi-definite bound both for the unitary states,
\be
\ba{ll}
\mbox{For~~}\Lambda(t)=0,~~~&~\dis{\dot{K}_{\mu}\dot{K}^{\mu}=4H(t)^{2}~\geq 0\,,}\\
{}&{}\\
\mbox{For~~}\Lambda(t)>0,~~~&~\dis{\dot{K}_{\mu}\dot{K}^{\mu}=8
\left\{\Q_{+}\Q_{+}^{\dagger}\,,\,\Q_{-}\Q_{-}^{\dagger}\right\}}
~\geq 0\,,\\
{}&{}\\
\mbox{For~~}\Lambda(t)<0,~~~&~\dis{\dot{K}_{\mu}\dot{K}^{\mu}=2
\left\{J_{+}\,,\,J_{+}^{\dagger}\right\}~\geq 0\,,}
\ea
\label{EVID3}
\ee
and also for the non-unitary states for which  $\Lambda>0$ and
$P$, $iX$ being Hermitian (or  anti-Hermitian),
\be
\dis{\dot{K}_{\mu}\dot{K}^{\mu}=\left[\trSU\!\left(P^{2}+\Lambda X^{2}\right)\right]^{2}-
\Lambda\left[\trSU\!\left(PX+XP\right)\right]^{2}\geq 0\,.}
\ee
The equality holds only for the trivial case, $X=P=0$. Thus,
the velocity vector, $\dot{K}_{\mu}$ is always space-like, which is natural for the Euclidean  geometry of the unitary states.  But in the Minkowskian space, as for  the non-unitary states with  $\Lambda>0$,  it implies  the superluminar behavior, i.e. ``tachyon".   As shown above, all the  BPS configurations  correspond to the null geometry, and hence not tachyonic. \\

To summarize, the normalizable or the unitary    sector  in  the Hilbert space relevant to the case   $\Lambda<0$ is characterized by  the Euclidean $dS_{2}/AdS_{2}$ geometry, while the non-normalizable or the non-unitary  sector relevant to the case $\Lambda>0$ has the  Minkowskian geometry.     All the BPS states  always  correspond to the null geometry, i.e. $\Cso=0$.   When $\Lambda(t)>0$ and  $\Cso<0$, i.e. the  Minkowskian \textit{de-Sitter} geometry, from (\ref{EVID3}), the particles  in the $\so(1,2)$ hyperspace are  tachyonic and can not be supersymmetric. \\

\subsection{Holographic dual to $2D$  superstring}
Various matrix models with  potentials having a single maximum have been proposed as dual candidates of $2D$ string theories on $AdS_{2}$-type backgrounds with the rolling tachyon or the linear dilaton~\cite{McGreevy:2003kb,Martinec:2003ka,Klebanov:2003km,Takayanagi:2003sm,
Douglas:2003up,Strominger:2003tm,Verlinde:2004gt}. The continuum or so called the double scaling limit in the matrix models zoom in on the maximum of the potential, effectively leaving a single upside down harmonic potential~\cite{Gross:1990ay,Brezin:1989ss,Ginsparg:1990as},  precisely  the same feature as our $\osp(1|2,\R)$   matrix model shares when $\Lambda>0$.  Furthermore, the Hermitian matrix itself is  supposed  to represent the non-Abelian open string tachyon~\cite{McGreevy:2003kb}, and this is manifest in   our dispersion relation, (\ref{EVID3}),   for the case  of $\Lambda>0$ and $\Cso<0$.  Thus, we conclude    that when $\Lambda(t)$ is positive,   the    $\osp(1|2,\R)$ $\M$-theory matrix model provides holographic duals  to the two dimensional  Minkowskian superstring theories. The relevant sector in the matrix model Hilbert space is  then  the non-normalizable or non-unitary  one  satisfying   $\Cso<0$.  From  (\ref{EVID3}),  the choice of the decreasing $\Lambda(t)$, like  $\Lambda(t)=e^{-t/t_{o}}$, seems   appropriate for   the description of   the tachyon condensation~\cite{SenTachyon} or the D-brane
decay~\cite{McGreevy:2003kb}. Further investigation is to be required.   \\

On the other hand, when  $\rho=0$,   for the constant positive  $\Lambda$  the generic BPS configurations  (\ref{BPSconf})  are given by the hyperbolic functions,
\be
\dis{X(t)=\cosh\left(\sqrt{{\Lambda}}\,t\right)X(0)+\frac{\kappa}{\sqrt{\Lambda}\,}
\sinh\left(\sqrt{\Lambda}\,t\right)1}\,,
\label{BPSTp}
\ee
while for the constant negative $\Lambda$  they are  the usual harmonic oscillators,
\be
\dis{X(t)=\cos\left(\sqrt{\left|{\Lambda}\right|}\,t\right)X(0)
+\frac{\kappa}{\sqrt{\left|{\Lambda}\right|}\,}
\sin\left(\sqrt{\left|{\Lambda}\right|}\,t\right)1}\,.
\label{BPSTn}
\ee
The latter is also consistent with the Euclidean $2D$ superstring theory or the $\N=2$ super Liouville theory  results~\cite{Eguchi:2003ik,Ahn:2003tt,Kutasov:2004dj,Ahn:2004qb,Nakayama:2004yx,Lapan:2005qz}. The classical shape of  the so called FZZT brane (falling Euclidean D0-brane),  which is given  as time dependent boundary state,  precisely matches with (\ref{BPSTn}). Thus, we expect  that when $\Lambda(t)$ is negative,  the   $\osp(1|2,\R)$ $\M$-theory matrix model provides holographic dual description of $2D$ Euclidean  superstring theories or superconformal theories. In particular, if $\Lambda$ is negative constant,  it corresponds to the  $\N=2$ super Liouville theory,  with the relation to the `Liouville background charge',   $Q_{\scriptscriptstyle{\,\rm Liouville}}=2\left|\Lambda\right|$.\\
~\\
~\\


\begin{center}
\large{\textbf{Acknowledgments}}
\end{center}
The author wishes to thank Xavier Bekaert, Nakwoo Kim,  Nikita Nekrasov, Gordon Semenoff,  Jan Troost,  Satoshi  Yamaguchi for valuable comments, and the organizers of  RTN Corfu Summer Institute $2005$ for the enlightening workshop where the present paper  was initiated. The work was partially supported by the European Research Training Network contract $005104$  "ForcesUniverse".

%

\end{document}